\documentclass[graybox]{svmult}

\usepackage{mathptmx}       % selects Times Roman as basic font
\usepackage{helvet}         % selects Helvetica as sans-serif font
\usepackage{courier}        % selects Courier as typewriter font
\usepackage{type1cm}        % activate if the above 3 fonts are
\usepackage{makeidx}         % allows index generation
\usepackage{graphicx}        % standard LaTeX graphics tool
\usepackage{multicol}        % used for the two-column index
\usepackage[bottom]{footmisc}% places footnotes at page bottom

\usepackage{amssymb}

\makeindex             % used for the subject index
\begin{document}

\title*{Magnetic field amplification in hypermassive neutron stars via the magnetorotational instability}
\titlerunning{Magnetic field amplification in HMNSs via the MRI}
\author{Daniel M. Siegel and Riccardo Ciolfi}
\institute{Daniel M. Siegel and Riccardo Ciolfi \at Max Planck Institute for
  Gravitational Physics (Albert Einstein Institute), Am M\"uhlenberg
  1, 14476 Potsdam-Golm, Germany} %, \email{daniel.siegel@aei.mpg.de}}
\maketitle

\abstract{Mergers of binary neutron stars likely lead to the formation
  of a hypermassive neutron star (HMNS), which is metastable and
  eventually collapses to a black hole. This merger scenario is
  thought to explain the phenomenology of short gamma-ray bursts
  (SGRBs). The very  high energies observed in SGRBs have been
  suggested to stem from neutrino-antineutrino annihilation and/or
  from very strong magnetic fields created during or after the merger
  by mechanisms like the magnetorotational instability (MRI). Here, we
  report on results that show for the first time the development of
  the MRI in HMNSs in three-dimensional,
  fully general-relativistic magnetohydrodynamic simulations. This
  instability amplifies magnetic fields exponentially and could be a
  vital ingredient in solving the SGRB puzzle.}

%%%%%%%%%%%%%%%%%%%%%%%%%%%%%%%%%%%%%%		
% INTRODUCTION
%%%%%%%%%%%%%%%%%%%%%%%%%%%%%%%%%%%%%%

\section{Introduction}

A significant fraction of neutron star -- neutron star (NS--NS) binary mergers 
can lead to the formation of a hypermassive neutron star (HMNS) that eventually
collapses to a stellar-mass black hole, surrounded by a hot and dense accretion
torus (e.g. \cite{Shibata06a, Rezzolla:2010}). Besides being among the
most promising sources for the first direct detection of gravitational
waves with advanced ground-based interferometers such as Advanced LIGO and VIRGO
\cite{Harry2010,Accadia2011}, the inspiral and coalescence of NS--NS binaries is also
thought to be the progenitor system for short gamma ray bursts
(SGRBs), the most luminous explosions observed in the universe (see, e.g.,
\cite{Piran:2004ba,Gehrels2009} for a review). The association of
SGRBs with NS--NS coalescence is supported on both observational
\cite{Barthelmy2005,Gehrels2005} and theoretical
\cite{Rezzolla:2011,Rosswog2013} grounds. The observed SGRB fluxes,
their cosmological distances and their duration of $< 2\,\mathrm{s}$
require highly relativistic motion with Lorentz factors of up to
several hundreds to resolve the so-called compactness problem
\cite{Piran:2004ba}, which states that in the absence of high Lorentz
factors, SGRBs would show a thermal spectrum --- in contradiction to the observed
non-thermal spectra. Apart from neutrino-antineutrino annihilation as
one possibility, the
enormous amounts of energy needed to generate such extreme velocities
have been suggested to stem from strong magnetic fields produced
during/after the merger process by mechanisms such as the Kelvin-Helmholtz
(KH) \cite{Price06,Anderson2008,Giacomazzo2013}
and the magnetorotational instability (MRI)
\cite{Duez:2006qe,Duez2006a}, or by high field strengths
generated in the torus after the central black hole has formed (as in
\cite{Rezzolla:2011}). One advantage of the former mechanisms is that
they do not depend on the creation and properties of a torus and that
they would already act prior to the collapse. While the amount of
amplification through the KH instability, triggered when the two stars
enter into contact, is controversial and maybe limited to only one order of magnitude
\cite{Price06,Giacomazzo2013}, the MRI triggered in
the metastable differentially rotating HMNS appears to constitute a
promising amplification mechanism, especially in the light of recent
results indicating a rather stiff equation of state
\cite{Demorest2010,Antoniadis2013}. The latter results suggest that HMNSs are
indeed a likely outcome of NS--NS mergers \cite{Hotokezaka2011,Hotokezaka2013}, and
that they are probably longer lived than previously thought, providing
more time for a potential MRI to act. However, simulating
the MRI in three dimensions under the extreme physical conditions of HMNS interiors is a
challenge and had not been accomplished until very recently \cite{Siegel2013}.

Here, we elaborate on these recent results that have shown for the first time
direct evidence for the MRI in the interior of a HMNS in global,
three-dimensional and fully general-relativistic magnetohydrodynamic
simulations.

%%%%%%%%%%%%%%%%%%%%%%%%%%%%%%%%%%%%%%		
% Setup
%%%%%%%%%%%%%%%%%%%%%%%%%%%%%%%%%%%%%%

\section{Capturing the MRI in HMNSs}

The magnetorotational instability \cite{Velikhov1959,Chandrasekhar1960}
can be triggered in differentially rotating magnetized
fluids \cite{Balbus1991} and refers to modes that grow
exponentially from initial seed perturbations. From a linear
perturbation analysis of the Newtonian MHD equations, one can estimate
the characteristic timescale $\tau_{_\mathrm{MRI}}$  and wavelength $\lambda_{_\mathrm{MRI}}$
for the fastest-growing mode with wave vector $\mathbf{k}$ by
\begin{equation}
\tau_{_\mathrm{MRI}} \sim \Omega^{-1}\,, \qquad
\lambda_{_\mathrm{MRI}} \sim
\left(\frac{2\pi}{\Omega}\right) 
\left(\frac{\mathbf{B\cdot}\mathbf{e}_\mathbf{k}}{\sqrt{4\pi\rho}}\right)\, ,
\label{eq:tau_lambda}
\end{equation}
where $\Omega$ denotes the angular velocity of the fluid, $\rho$ the
density, $\mathbf{B}$ the magnetic field, and $\mathbf{e}_\mathbf{k}$
the unit vector in direction of $\mathbf{k}$
\cite{Balbus1991,Balbus1995}. We note that there is no general
analytic description of the MRI within general-relativistic MHD to
date. Nevertheless, we can compare our fully general-relativistic
numerical results with the above analytical estimates, provided that the
Newtonian predictions are translated into general relativity by
employing equivalence principle arguments \cite{Siegel2013}.

Resolving the MRI in an MHD simulation is a challenge, as
$\lambda_{_\mathrm{MRI}}$ is typically much smaller than the characteristic
length scale of the astrophysical system under study. Local
simulations of only a small part of the system
(e.g. \cite{Obergaulinger:2009}) or simulations in axisymmetry
\cite{Duez:2006qe,Duez2006a} are usually
conducted in order to render the problem computationally
affordable. In the case of HMNSs, capturing
the MRI for realistic (i.e. relatively low) magnetic field strengths is particularly
demanding due to the extremely high densities and high angular
velocities involved (cf. Eq. (\ref{eq:tau_lambda})).

Here, we discuss global three-dimensional simulations, which start
from a typical axisymmetric and differentially rotating HMNS model of mass
$M=2.23$\,M$_\odot$ and central angular velocity $\Omega_c=2\pi\times
7\,\mathrm{kHz}$ (representing the outcome of a NS--NS merger) and
employ a four-level nested-boxes grid hierarchy, with the finest refinement
level covering the HMNS at all times. In order to capture the MRI and
despite the very high resolutions used here (with finest grid spacing of
$h=44\,\mathrm{m}$), high initial magnetic field strengths of
$B^{\mathrm{in}}=(1-5)\times 10^{17}\,\mathrm{G}$ have to be employed,
assuming that these field strengths have previously been generated by
compression during the merger, the KH instability, magnetic winding
and previous MRI activity. However, it is important to point out that
these field strengths are still very small in terms of the average
magnetic-to-fluid pressure ratio, which is between $0.045-1.2\times
10^{-2}$. In order to reduce the computational costs, a reflection
symmetry across the $z=0$ plane and a $\pi/2$ rotation
symmetry around the $z$-axis have been applied. By performing two
additional simulations, removing either the reflection symmetry or
replacing the rotation symmetry by a $\pi$ symmetry, we have verified
that these discrete symmetries do not significantly influence our
results. For instance, the relative differences for the maximum of the
toroidal field strength (as plotted in Fig.~\ref{fig:box_analysis})
are well below $10^{-3}$ up to $t\approx 0.4\,\mathrm{ms}$ when the
star starts collapsing to a black hole.

%%%%%%%%%%%%%%%%%%%%%%%%%%%%%%%%%%%%%%
% RESULTS
%%%%%%%%%%%%%%%%%%%%%%%%%%%%%%%%%%%%%%

\section{Discussion of simulation results}

\begin{figure}[t]
  \includegraphics[width=\textwidth]{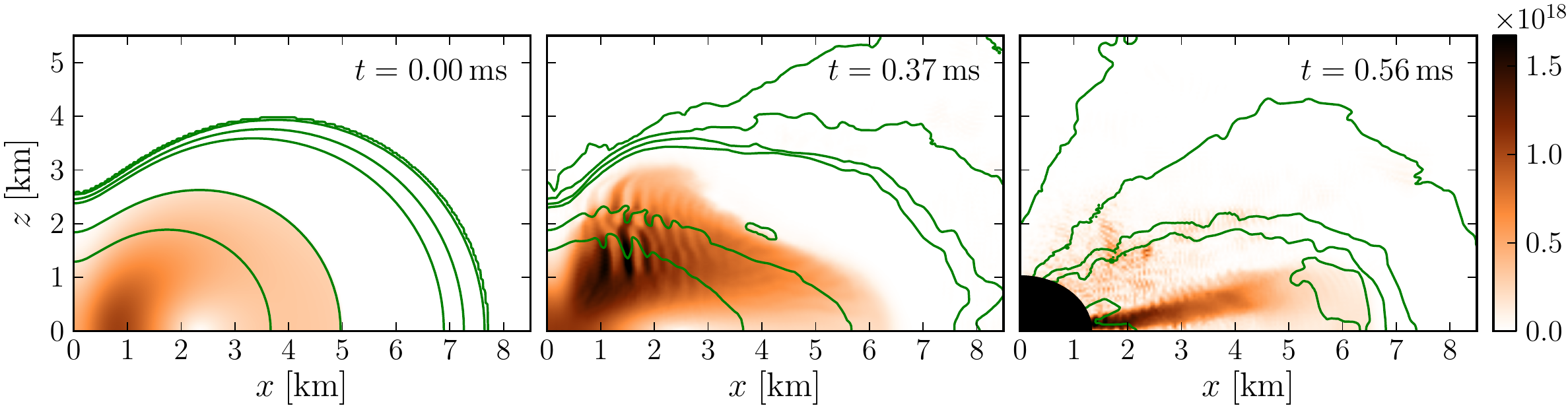}
  \caption{Three characteristic stages of the HMNS evolution represented by a cut in
    the $x$-$z$ plane, showing the colour-coded norm of the magnetic field
    (in G) and selected density contours: initial HMNS model,
    pronounced MRI development, and early post-collapse phase with a black hole
    (horizon is masked) surrounded by a magnetized torus.}
\label{fig:evolution}
\end{figure}

\begin{figure}[t]
  \centering
  \includegraphics[width=0.48\textwidth]{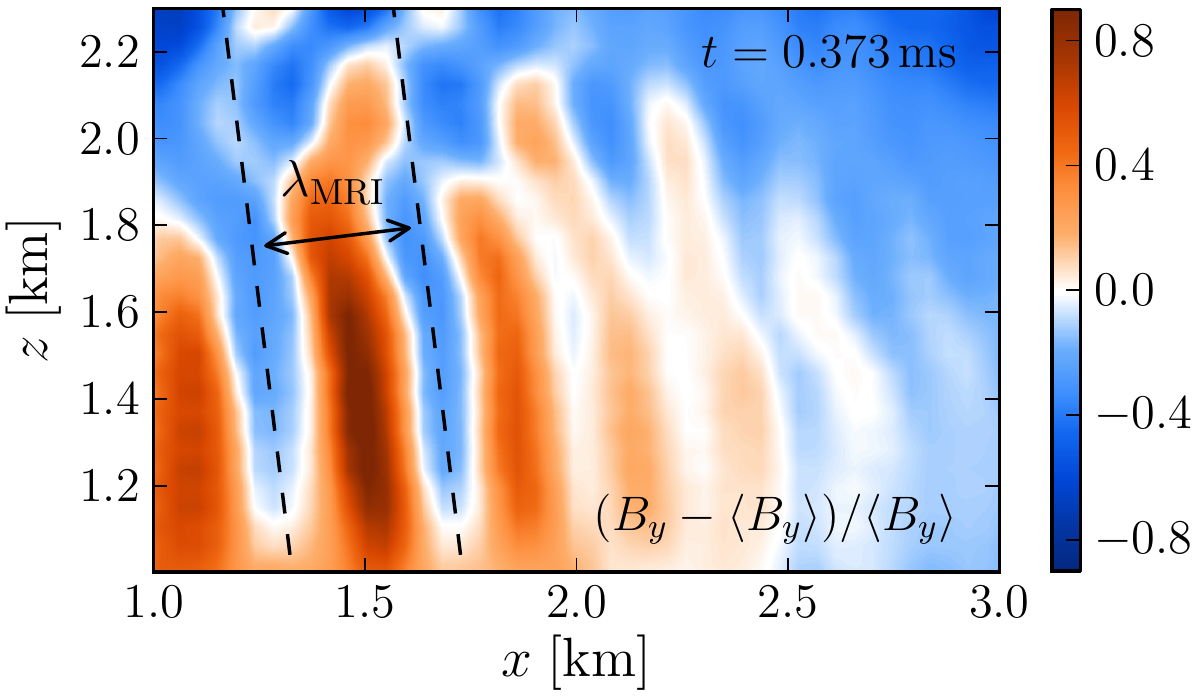}
  \hspace{0.2cm}
  \includegraphics[width=0.48\textwidth]{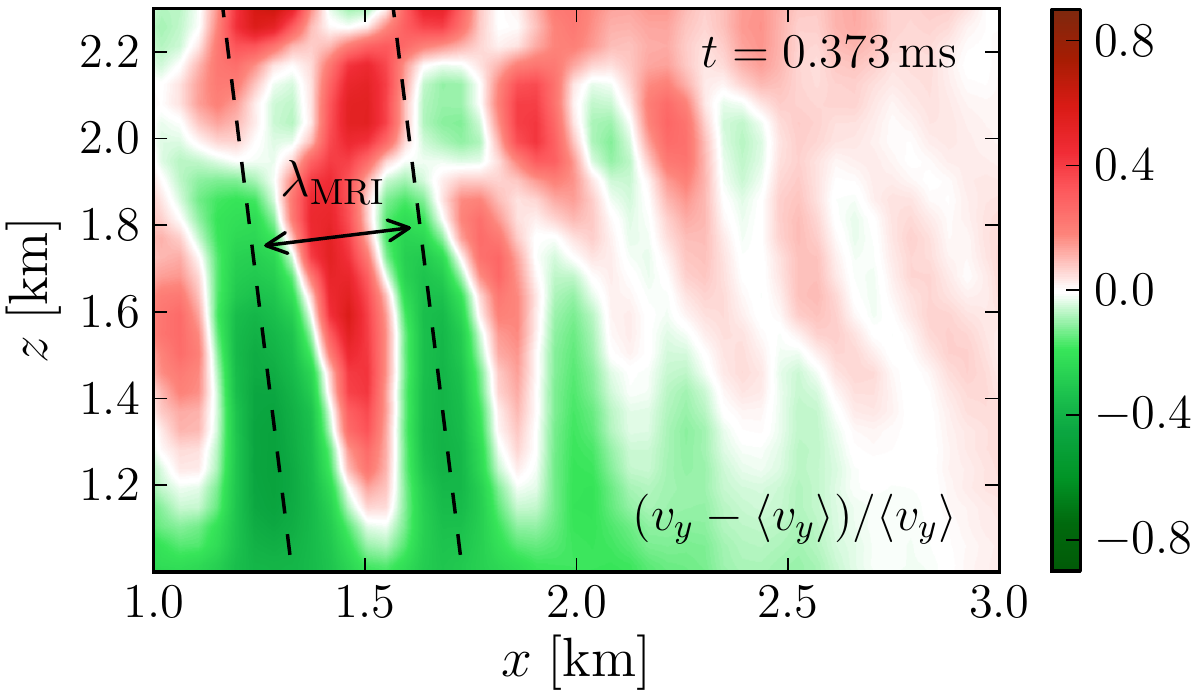}
  \caption{Channel flow structures as seen in the toroidal magnetic
    field (left) and the toroidal component of the 3-velocity field
    (right). Shown are the relative deviations from the mean value in the
    region considered here. The wavelength of the fastest-growing MRI mode
    can be clearly identified.}
\label{fig:channel_flows}
\end{figure}

Figure \ref{fig:evolution} provides a representative overview of the
HMNS evolution: the initial axisymmetric configuration, which
shows a highly flattened HMNS due to rapid rotation; the stage
of a developed MRI, indicated by the ripples in the
magnetic field and density; the early post-collapse phase showing a
black hole surrounded by a magnetized and geometrically thick
torus. The ``ripple patterns'' seen in this simulation are similar to the coherent
channel flow structures (WKB-like modes) observed in local axisymmetric Newtonian MRI
simulations, which are the characteristic signatures of this
instability (e.g. \cite{Obergaulinger:2009}). This is the first time
in global general-relativistic simulations of HMNSs that such
rapidly-growing and spatially-periodic structures are observed. Figure
\ref{fig:channel_flows} displays the channel flow structures in a
zoomed-in version for the toroidal magnetic field and the toroidal
component of the velocity field. The onset of channel-flow merging
(reminiscent of the results reported in \cite{Obergaulinger:2009}) is
evident in the upper part of the panels.

\begin{figure}[t]
  \centering
  \includegraphics[width=0.48\textwidth]{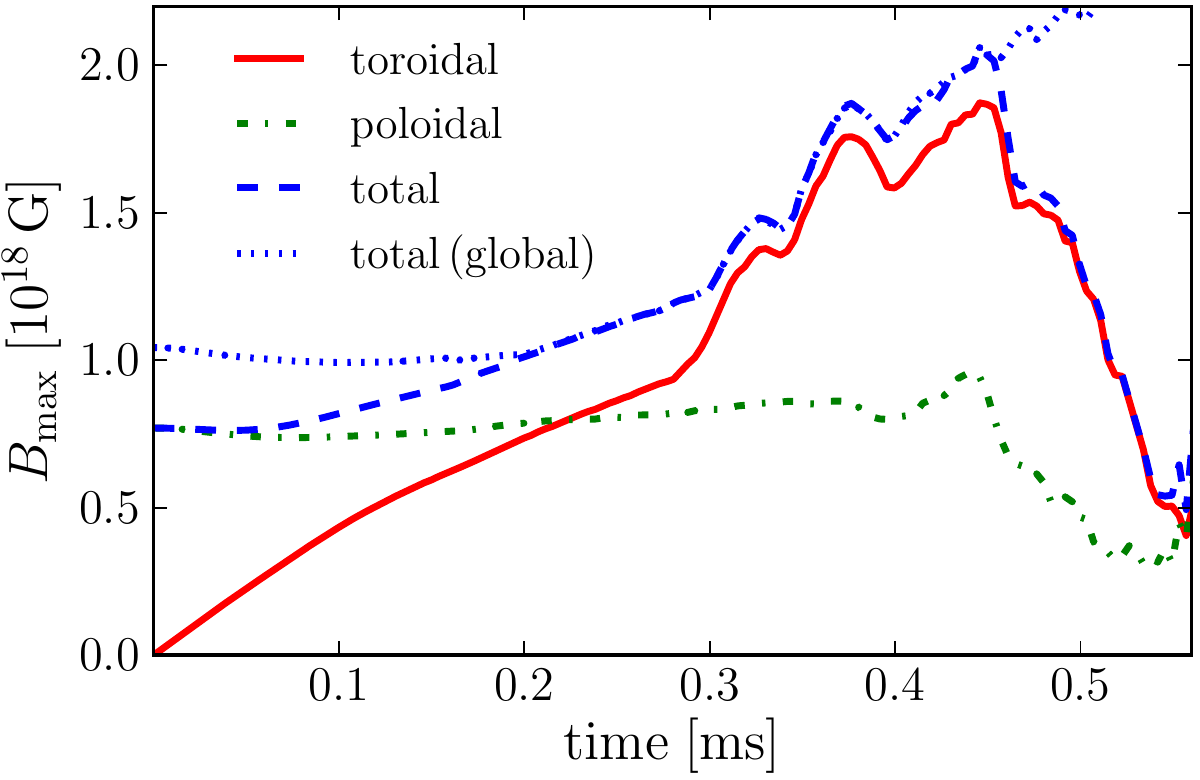}
  \hspace{0.2cm}
  \includegraphics[width=0.48\textwidth]{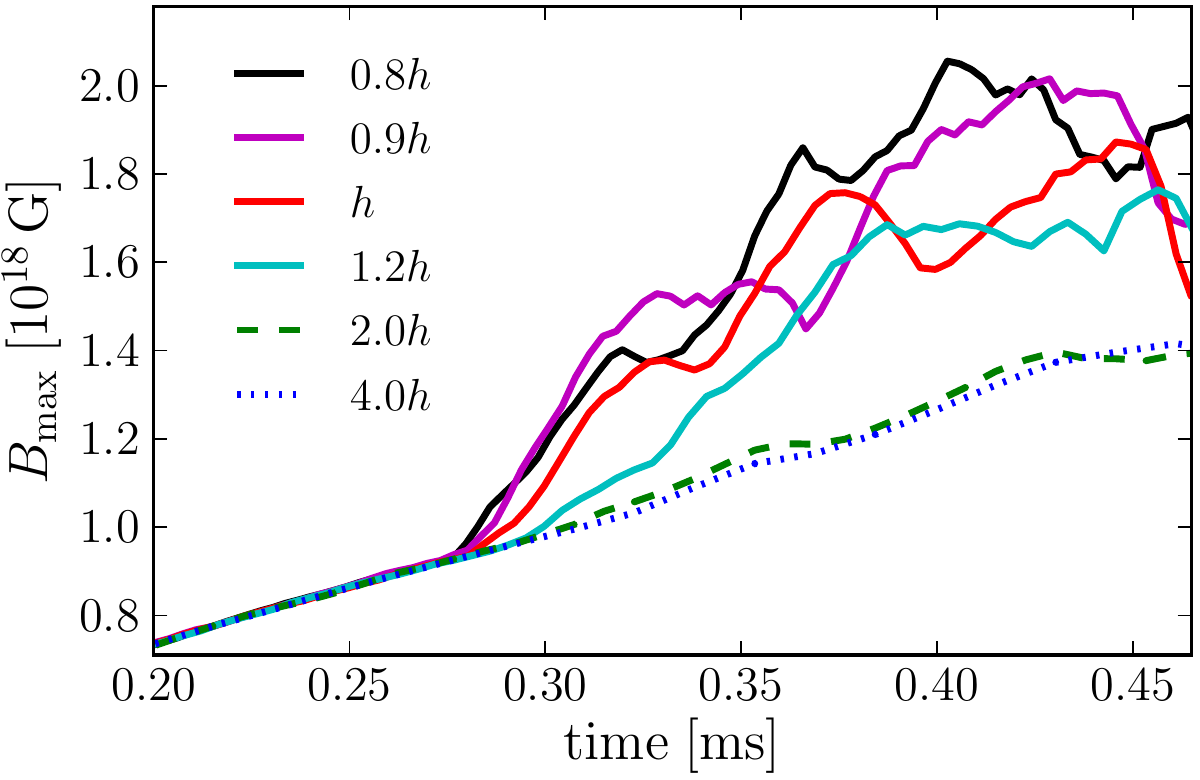}
  \caption{Left: maximum magnetic field strength for the toroidal,
    poloidal and total magnetic field in the region
    $(x,z)=[1.0,3.0]\times [1.0,2.3]\,\mathrm{km}$, and the
    corresponding total evaluated in the entire $x$-$z$ plane. Right:
    Maximum toroidal magnetic field strengths in the same region for
    runs with different resolution (but otherwise identical), with
    finest grid spacings between $(4-0.8)h$, where $h=44\,\mathrm{m}$
    refers to the fiducial value.}
\label{fig:box_analysis}
\end{figure}

The left panel of Fig.~\ref{fig:box_analysis} shows the maximum
field strength for the toroidal, poloidal and total magnetic field in
the region $(x,z)\in [1.0,3.0]\times [1.0,2.3]\,\mathrm{km}$, where the instability
develops most prominently, along with the analogous quantity for the
total field evaluated over the entire $x$-$z$ plane. The purely
poloidal initial seed field geometry is lost very soon due to magnetic winding, which
leads to a linear increase in the toroidal field strength. At
$t\approx 0.3\,\mathrm{ms}$, exponential magnetic field amplification due to
the MRI sets in and lasts up to $t\approx 0.4\,\mathrm{ms}$ when the
star starts collapsing to a black hole. It is important to point out
that even during this very short time frame, the MRI contributes
significantly to a global magnetic field amplification of the system
(compare local with global maximum of the total field). 

The right
panel of Fig.~\ref{fig:box_analysis} demonstrates that the onset of
the instability is well resolved. While any sign of the instability is
absent in the case where there are less than five grid points per
wavelength of the fastest-growing mode (as given by
Eq.~(\ref{eq:tau_lambda})), we gradually recover the
growth rate of the fastest-growing mode with increasing
resolution. The extracted values for this growth rate agree within
error bars among the three highest-resolution runs, and the resulting
value of $\tau_\mathrm{MRI}=(8.2\pm 0.4)\times 10^{-2}\,\mathrm{ms}$
is in order-of-magnitude agreement with the Newtonian analytic
prediction from Eq.~(\ref{eq:tau_lambda}) for the considered region
once translated to our general-relativistic setting,
$\tau_\mathrm{MRI}=(4-5)\times 10^{-2}\,\mathrm{ms}$. Also the
wavelength of the fastest-growing mode as measured with a Fourier
analysis of the magnetic field in the selected region
($\lambda_\mathrm{MRI} \approx 0.4\,\mathrm{km}$) is in
order-of-magnitude agreement with the corrected Newtonian analytic
prediction from Eq.~(\ref{eq:tau_lambda}) for this region,
$\lambda_\mathrm{MRI}\approx (0.5-1.5)\,\mathrm{km}$. For further
details and verification of additional properties of the MRI as expected
from local Newtonian simulations in other astrophysical systems, we
refer to \cite{Siegel2013}.

%%%%%%%%%%%%%%%%%%%%%%%%%%%%%%%%%%%%%%
% CONCLUSIONS
%%%%%%%%%%%%%%%%%%%%%%%%%%%%%%%%%%%%%%

\section{Conclusion}

The simulations discussed here have shown for the first time direct
evidence for the MRI in HMNSs in three-dimensional, fully
general-relativistic MHD simulations. This evidence manifests itself, e.g.,
in the presence of coherent channel flow structures, which have not 
been previously observed in three-dimensional general-relativistic
MHD simulations. The growth of these structures coincides with
exponential growth in the toroidal field strength. Furthermore, the
two characteristic quantities of the instability, the growth time and
wavelength of the fastest-growing mode, could be measured directly, and
the resulting values are in order-of-magnitude agreement with the
simplified Newtonian analytical estimates, once the latter
have been corrected for coordinate effects due to the general
relativistic framework of the simulations. 

It is interesting to note that these simulations also represent the
first detailed observation of the MRI in the strong gravity regime,
where the characteristic length scale of spacetime curvature becomes
comparable to the wavelength of the fastest-growing MRI mode. Despite
the fact that the existence of WKB-type modes becomes less obvious in
this context, these modes are observed and the idealized
Newtonian analytic description still seems to provide reasonable
predictions.

The global magnetic field amplification due to the MRI leads 
to very strong magnetic fields that -- at least for some time -- stay in the
vicinity of the newly formed black hole, and can thus potentially contribute
to power a relativistic jet launched by the black-hole-torus
system. Therefore, the presence of the MRI in HMNSs is of great astrophysical
importance, as it could be a vital ingredient in solving the
compactness problem of SGRBs. 

One open question is to understand how 
much amplification can be achieved in a HMNS, before the dominant saturation 
mechanisms take place. In our model, the relatively short life of the HMNS
limits the amplification to less than one order of magnitude. At the time of the 
collapse there is still no sign of saturation, which suggests the
possibility of much higher magnetic field amplification in
longer-lived models. This will be the focus of future studies.

%%%%%%%%%%%%%%%%%%%%%%%%%%%%%%%%%%%%%%
% ACKNOWLEDGMENTS
%%%%%%%%%%%%%%%%%%%%%%%%%%%%%%%%%%%%%%

\section*{Acknowledgments}

DMS greatly acknowledges the award of the first
Karl Schwarzschild Prize sponsored by Springer for the best talk in
the student section of the first
Karl Schwarzschild Meeting, held in Frankfurt, Germany, July 2013. DMS
also thanks the organizers of this meeting for travel support.

%%%%%%%%%%%%%%%%%%%%%%%%%%%%%%%%%%%%%%
%%%%%%%%%%%%%%%%%%%%%%%% referenc.tex %%%%%%%%%%%%%%%%%%%%%%%%%%%%%%
% sample references
% %
% Use this file as a template for your own input.
%
%%%%%%%%%%%%%%%%%%%%%%%% Springer-Verlag %%%%%%%%%%%%%%%%%%%%%%%%%%
%
% BibTeX users please use
 \bibliographystyle{spphys}
 \bibliography{aeireferences}

\begin{thebibliography}{10}
\providecommand{\url}[1]{{#1}}
\providecommand{\urlprefix}{URL }
\expandafter\ifx\csname urlstyle\endcsname\relax
  \providecommand{\doi}[1]{DOI \discretionary{}{}{}#1}\else
  \providecommand{\doi}{DOI \discretionary{}{}{}\begingroup
  \urlstyle{rm}\Url}\fi

\bibitem{Shibata06a}
M.~{Shibata}, K.~{Taniguchi}, Phys. Rev. D \textbf{73}, {064027} (2006)

\bibitem{Rezzolla:2010}
L.~{Rezzolla}, L.~{Baiotti}, B.~{Giacomazzo}, et~al., Class.
  Quantum Grav. \textbf{27}(11), 114105 (2010)

\bibitem{Harry2010}
G.M.~{Harry}, et~al., Class. Quantum Grav. \textbf{27}, 084006 (2010)

\bibitem{Accadia2011}
T.~{Accadia}, F.~{Acernese}, F.~{Antonucci}, et~al.,
Class. Quantum Grav. \textbf{28}(11), 114002 (2011)

\bibitem{Piran:2004ba}
T.~Piran, Rev. Mod. Phys. \textbf{76}, 1143 (2004)

\bibitem{Gehrels2009}
N.~{Gehrels}, E.~{Ramirez-Ruiz}, D.B.~{Fox}, Ann. Rev. Astron. \&
  Astrophys. \textbf{47}, 567 (2009)

\bibitem{Barthelmy2005}
S.D. {Barthelmy}, G.~{Chincarini}, D.N.~{Burrows}, et~al., Nature \textbf{438}, 994 (2005)

\bibitem{Gehrels2005}
N.~{Gehrels}, C.L.~{Sarazin}, P.T.~{O'Brien}, et~al., Nature \textbf{437}, 851 (2005)

\bibitem{Rezzolla:2011}
L.~{Rezzolla}, B.~{Giacomazzo}, L.~{Baiotti}, et~al., Astrophys. J. \textbf{732}(11), L6 (2011)

\bibitem{Rosswog2013}
S.~{Rosswog}, T.~{Piran}, E.~{Nakar}, Mon. Not. R. Astron. Soc. \textbf{430}, 2585 (2013)

\bibitem{Price06}
R.H.~{Price}, S.~{Rosswog}, Science \textbf{312}, 719 (2006)

\bibitem{Anderson2008}
M.~{Anderson}, E.W.~{Hirschmann}, L.~{Lehner}, et~al., Phys. Rev. Lett. \textbf{100},
  191101 (2008)

\bibitem{Giacomazzo2013}
B.~{Giacomazzo}, R.~{Perna}, Astrophys. J. Lett. \textbf{771}, L26 (2013)

\bibitem{Duez:2006qe}
M.D.~{Duez}, Y.T.~{Liu}, S.L.~{Shapiro}, et~al., Phys. Rev. D
  \textbf{73}, 104015 (2006)

\bibitem{Duez2006a}
M.D.~{Duez}, Y.T.~{Liu}, S.L.~{Shapiro}, et~al., Phys.
  Rev. Lett. \textbf{96}(3), 031101 (2006)

\bibitem{Demorest2010}
P.B.~{Demorest}, T.~{Pennucci}, S.M.~{Ransom}, et~al., Nature \textbf{467}, 1081 (2010)

\bibitem{Antoniadis2013}
J.~{Antoniadis}, P.C.C.~{Freire}, N.~{Wex}, et~al., Science \textbf{340}, 448 (2013)

\bibitem{Hotokezaka2011}
K.~{Hotokezaka}, K.~{Kyutoku}, H.~{Okawa}, et~al., Phys.
  Rev. D \textbf{83}(12), 124008 (2011)

\bibitem{Hotokezaka2013}
K.~{Hotokezaka}, K.~{Kiuchi}, K.~{Kyutoku}, et~al., Phys. Rev. D \textbf{87}(2), 024001 (2013)

\bibitem{Siegel2013}
D.M.~{Siegel}, R.~{Ciolfi}, A.I.~{Harte}, et~al., Phys. Rev. D R
  \textbf{87}(12), 121302 (2013)

\bibitem{Velikhov1959}
E.P.~{Velikhov}, Sov. Phys. JETP \textbf{36}, 995 (1959)

\bibitem{Chandrasekhar1960}
S.~{Chandrasekhar}, Proc. Natl. Acad. Sci. \textbf{46}, 253 (1960)

\bibitem{Balbus1991}
S.A.~{Balbus}, J.F.~{Hawley}, Astrophys. J. \textbf{376}, 214 (1991)

\bibitem{Balbus1995}
S.A.~{Balbus}, Astrophys. J. \textbf{453}, 380 (1995)

\bibitem{Obergaulinger:2009}
M.~{Obergaulinger}, P.~{Cerd{\'a}-Dur{\'a}n}, E.~{M{\"u}ller}, et~al.,
  Astron. Astrophys. \textbf{498}, 241 (2009)

\end{thebibliography}

\end{document}